\documentstyle[preprint,aps,psfig]{revtex}

\begin{document} 

\title{Reversed anisotropies and thermal
contraction of FCC (110) surfaces}

\author{Shobhana Narasimhan}

\address{ Theoretical Sciences Unit, Jawaharlal Nehru Centre for
Advanced Scientific Research,\\ Jakkur PO, Bangalore 560 064, India}

\date{August 3 2000}

\maketitle

\begin{abstract}

The observed anisotropies of surface vibrations for unreconstructed FCC
metal (110) surfaces are often reversed from the ``common sense"
expectation.  The source of these reversals is investigated by
performing {\em ab initio} density functional theory calculations to
obtain the surface force constant tensors for Ag(110), Cu(110) and
Al(110). The most striking result is a large enhancement in the
coupling between the first and third layers of the relaxed surface,
which strongly reduces the amplitude of out-of-plane vibrations of
atoms in the first layer. This also provides a simple explanation for
the thermal contraction of interlayer distances. Both the anisotropies
and the thermal contraction arise primarily as a result of the bond
topology, with all three (110) surfaces showing similar behavior.

\end{abstract}

\newpage 
The low-index surfaces of face-centered cubic (FCC) metals are
arguably the most studied of surfaces. Though they are often intended
to serve as prototypes for understanding the behavior of more complex
surface systems, it has become evident in the last two decades that
even these ``simple" surfaces display a wide range of complicated and
often counter-intuitive phenomena.  In this paper, I consider the case
of unreconstructed FCC (110) surfaces.  I try to show that some of the
observed features of their thermal behavior that have generally been
accepted as violating ``common sense" can actually be explained by
simple arguments, backed up with results from {\em ab initio} density
functional theory calculations.  These results also have important
implications for the study of other surface phenomena, and can, for
example, provide insight into the mechanisms behind
temperature-dependent surface phase transitions such as roughening and
premelting.

More bonds are broken to create a (110) surface from a bulk FCC
crystal, than for a (100) or (111) surface, and one therefore expects
the departures from bulk-like behavior to be largest for a (110)
surface. Structurally, this is manifested in FCC metals in one of two
ways: either the surface reconstructs into a ``missing row" structure
\cite{misrow}, or the surface unit cell remains unchanged but there is
a very pronounced oscillatory relaxation of interlayer spacings near
the surface \cite{oscill}. In either case, the very existence of the
surface lowers symmetry, and one expects to find anisotropic
modifications in the thermal properties of atoms at or near the
surface.

The surprise is that for these unreconstructed FCC(110) surfaces, some
of these anisotropies are {\em reversed}, i.e., they violate the
``common sense" expectation. For example, experiments show that in
general:  (i) atoms in the topmost surface layer have a bigger
mean-squared displacement (MSD) in the surface plane than normal to it,
whereas one might have expected the latter to be larger, since along
this direction, there are no higher layers to repel the atoms in the
first layer \cite{agmeis,almeis}; (ii) the MSDs normal to the surface
are larger for second layer atoms than for first layer atoms, though
one would expect that the MSDs should decay monotonically into the
bulk\cite{alleed1};  (iii) a third curious fact about unreconstructed
FCC (110) surfaces is that in some cases there is a thermal {\em
contraction} of interlayer distances upon heating.  The rule of thumb
seems to be that if the first interlayer spacing $d_{12}$ expands upon
heating, then the next interlayer spacing $d_{23}$ contracts
\cite{agmeis}; however, if $d_{12}$ exhibits a thermal contraction,
then $d_{23}$ expands with increasing
temperature\cite{alleed1,alleed2}.

Some of these patterns of behavior have also been observed in molecular
dynamics (MD) simulations \cite{talat,nicu110,aimd} but their origin
remains a puzzle.  In order to gain physical insight, and to be able to
predict trends across differently oriented surfaces and different
elements, one would like to not only reproduce this behavior in a
theoretical calculation, but also to know where these anomalous
properties arise from.  Are they primarily a consequence of the bond
topology, or are they due to peculiarities in the electronic
structure?  Such questions can reliably be answered by breaking up the
process of obtaining a fully relaxed surface from the cleavage of a
bulk crystal into steps, and performing a sequence of {\em ab initio}
calculations at each component step.

There have been few {\it ab initio} calculations of the lattice
dynamics and thermal behavior of the (110) surfaces of FCC metals.
Most of the previous calculations have been empirical (involving {\it
ad hoc} modifications of near-surface interactions so as to reproduce
experimental phonon spectra)\cite{stroscio,persson,black,franchini1},
or semi-empirical \cite{talat,nicu110}. The problem is, of course, that
this procedure is not guaranteed to give a unique result , and one may
be fooled into making modifications that are actually very far from the
correct ones. The only FCC(110) surface that has been investigated in
detail, using first-principles methods, is the Al(110) surface. The
pioneering calculations of Ho and Bohnen\cite{handb} and other
researchers \cite{gaspar,franchini2} of the lattice dynamics of Al(110)
have recently been followed by an {\em ab initio} MD study \cite{aimd},
which reproduces the experimentally observed anisotropies and the
thermal contraction of $d_{12}$.  But can the behavior of a simple
metal like Al also serve as an adequate paradigm for transition or
noble metals?  Computational constraints make it currently unfeasible
to perform such {\it ab initio} MD simulations on the noble metals or
on transition metals.

However, {\em ab initio} investigations of vibrational properties using
the ``frozen phonon" approach are possible, and have in fact been
performed by previous authors on other low-index surfaces of these
materials \cite{ho}. In this paper, I present the results of a series
of such frozen phonon calculations on the bulk FCC structures as well
as the unreconstructed (110) surfaces of Ag, Cu and Al.

For each case, I have determined the exact interlayer and/or intralayer
force constants from first principles, by distorting the equilibrium
structure and computing the Hellman-Feynman forces \cite{helfeyn}
thereby induced on the atoms. These forces, as well as the relaxed
structures, were obtained by performing density functional theory
calculations using the package ``fhi96md" \cite{fhi96md}, with fully
separable norm-conserving pseudopotentials \cite{tmpp} and  the
Ceperley-Alder form of the local density approximation \cite{cepaldxc}.
The calculations were carried out using plane wave basis sets with
energy cut-offs of 60 Ry, 70 Ry and 20 Ry for Ag, Cu and Al
respectively. It has been verified in earlier studies that the
pseudopotentials and basis sets used here work well for both bulk and
surface properties of these metals \cite{mine}. The surface
calculations for Ag and Cu were carried out using a repeating slab
geometry comprised of seven layers of atoms, separated by a vacuum
thickness equivalent to five layers; while a 15-layer slab was used for
the calculations on Al(110).  Bulk calculations were carried out using
both the conventional cubic unit cell and a unit cell containing 12 to
16 layers of atoms stacked along the [110] direction.

The interlayer force constants $\phi_{\alpha\beta}(i\ j)$ were obtained
by moving the atoms in a layer $i$ along the direction $\alpha$, and
computing the force along $\beta$ then experienced by atoms in other
layers $j$. For the case of bulk layers stacked along [110], I find
that for all three materials, the only significant elements of the
interlayer force constant tensors are the diagonal terms $\phi_{xx}(i,
\pm 1)$, $\phi_{yy}(i,i\pm 1)$ and $\phi_{zz}(i,i\pm 1)$ coupling
nearest-neighbor layers, and $\phi_{zz}(i,i\pm 2)$ which couples
next-nearest-neighbor layers (with the Cartesian directions defined as
$\hat x=[1\bar10], \hat y=[001]$ and $\hat z=[110]$). All other
elements of the interlayer force constant tensors are either zero by
symmetry, or are smaller by at least an order of magnitude. Similarly,
for the  relaxed surfaces, the only significant terms involving the
first (topmost) layer of atoms are $\phi_{xx}(1\ 2)$,
$\phi_{yy}(1\ 2)$, $\phi_{zz}(1\ 2)$ and $\phi_{zz}(1\ 3)$.

Some of these results are presented in Figure 1. For all three
materials, I find that the elements of the interlayer force constant
tensors stiffen considerably upon going from the bulk to the relaxed
surface.  The most notable feature is a dramatic increase in
$\phi_{zz}(1\ 3)$, whose magnitude is almost doubled relative to the
bulk value $\phi_{zz}(i,i\pm 2)$. In fact, $\phi_{zz}(1\ 3)$ is found
to be significantly larger than $\phi_{zz}(1\ 2)$.  This means that if
the atoms in layer 1 are displaced along the $z$ direction (normal to
the surface) the resulting force along $z$ experienced by atoms in
layer 3 is considerably greater than the force on atoms in layer 2,
which is a surprising and counter-intuitive result. Though the result
$\phi_{zz}(1\ 3) > \phi_{zz}(1\ 2)$ was also obtained by some previous
authors who fit force constant models to empirical data for
Al(110)\cite{persson} and Ni(110)\cite{nilehwald},  its importance
seems to have been overlooked. I will argue that this large enhancement
in $\phi_{zz}(1\ 3)$ is largely responsible for the anomalous thermal
behavior of these surfaces.

It should be emphasized that the results I have obtained for interlayer
force constant tensors are exact, and do not involve any assumptions
about the form or range of interatomic potentials. However, I now map
these results onto a model potential, in order to better understand the
implications of the changes in the surface force constant tensors.

For each case, I have first assembled an extremely large database of
results from {\em ab initio} frozen-phonon calculations. In addition to
the terms listed above, the database includes results for other
interlayer force constants, as well as intralayer terms. This database
is then used to parametrize a simple form of interatomic potential: for
each pair of nearest-neighbor (NN) atoms the interatomic potential
$U(r)$ is specified by a tangential parameter
$\alpha=r_0^{-1}[dU(r)/dr]_{r=r_0}$ and a radial parameter
$\beta=[d^2U(r)/dr^2]_{r=r_0}$, where $r_0$ is the equilibrium value of
the interatomic distance $r$.

For the interaction between two NN atoms in the bulk, I obtain
$\{\alpha_{bb}$,$\beta_{bb}\}$ = $\{-.0007, .0181\}, \{-.0006, .0236\}$
and $\{-.0007, .0152\}$ for Ag, Cu and Al respectively, with all force
constants being expressed in atomic units $(Ha/bohr^2)$.  The
oscillatory relaxation and charge redistribution at the relaxed
surfaces result in a modification of these values.  Accordingly, four
kinds of NN bonds $i$-$j$  between atoms in near-surface layers $i$ and
$j$ (1-1, 1-2, 1-3 and 2-3), are described by new parameters
$\alpha_{ij}$ and $\beta_{ij}$, while all other terms are left
unchanged from the bulk values.

Though this form of potential is admittedly simple, it is able to
reproduce the gross details of the bulk phonon spectrum, and suffices
to bring out the essential physics behind the altered surface behavior.
Moreover, though the parameter set is heavily overdetermined (with the
eight modified surface parameters being fit to a database of 34
different numbers determined from {\em ab initio} calculations for 13
different kinds of surface distortions, supplemented by three stability
criteria), the quality of the fit is reasonable for Al, and extremely
good for Ag and Cu.

Next, the model potential is used to set up the dynamical matrix for a
slab composed of many layers $N$ stacked along [110], which is then
diagonalized to obtain phonon frequencies $\omega_{{\bf k}\lambda}$ and
eigenvectors $e_{i\alpha}^{{\bf k}\lambda} $. The $\alpha$th component
of the MSDs at temperature $T$ for atoms in layer $i$, $\langle
u_{i\alpha}^2 (T)\rangle$, is then given by \cite{msdform}:

\begin{equation}\label{eq:msd} \langle u_{i\alpha}^2 (T)\rangle =
{1\over N}
		       \sum_{{\bf k}\lambda} {\hbar\over M\omega_{{\bf
		       k}\lambda}} (e_{i\alpha}^{{\bf k}\lambda} )^2
		       (n_{{\bf k}\lambda} + \textstyle{1\over2});
\end{equation} where the  summation runs  over all wave-vectors {\bf k}
in the surface Brillouin zone and all phonon branches $\lambda$;
$\hbar$ is Planck's constant, $M$ is the atomic mass and $n_{{\bf
k}\lambda}$ is the Bose-Einstein distribution factor.

In order to disentangle geometric effects from electronic ones, for
each material I consider various cases.  First, to determine
the consequences of the reduced co-ordination at the surface {\em
alone}, all NN interactions are replaced  by the bulk parameters
$\alpha_{bb}$ and $\beta_{bb}$. Upon using Equation (1), I find that
all components of the MSDs of atoms in the first two layers are larger
than the corresponding bulk values, but the anisotropies differ from
those seen in experiments and MD simulations:  the largest enhancement
is in $\langle u_{1y}^2\rangle$ and $\langle u_{1z}^2\rangle$, which
are both approximately 2.2 times as large as in the bulk, and $\langle
u_{1z}^2\rangle > \langle u_{2z}^2\rangle$.  Hitherto, there has been a
tendency to attribute any anomalies in the behavior of FCC(110)
surfaces to the ``very open" surface structure. However, these results
show that the open structure alone is not sufficient to explain the
observed phenomena.

The situation is considerably altered for the three relaxed surfaces.
The changes in the surface force constant tensors result in a
considerable increase (15 to 50 \%) in the value of the radial term
$\beta_{12}$, and an even larger increase (45 to 85\%) in $\beta_{13}$,
relative to $\beta_{bb}$. The huge enhancement in $\beta_{13}$
corresponds to the very large value obtained for $\phi_{zz}(1\ 3)$, and
implies that the bonds between NN atoms in layers 1 and 3 are extremely
stiff. However, I find that the radial term coupling two NN surface
atoms, $\beta_{11}$, is softened by $\sim$20\%. These results differ
considerably from the 40\% softening suggested for $\beta_{11}$  for
Ag(110) by  Franchini {\em et al.} \cite{franchini1}, and the softening
of $\beta_{12}$ by 6\% for Cu(110) suggested by Black {\em et
al.}\cite{black}, who fit the parameters of their models to
experimentally measured phonon spectra.

What is responsible for the increased stiffness of the 1-3 NN bonds? To
answer this, I looked also at the intermediate case of bulk-truncated
surfaces (with electronic relaxation permitted, but all interlayer
distances set equal to the bulk value).  I find that though the value
of $\beta_{13}$ relative to $\beta_{bb}$ is slightly modified for the
bulk-truncated surfaces, the huge enhancement comes upon going from the
bulk-truncated surface to the fully relaxed one.

The enhancement in $\beta_{13}$ is sufficiently large to push $\langle
u_{1z}^2\rangle$ down significantly, now making $\langle
u_{1z}^2\rangle < \langle u_{2z}^2\rangle $
 and  $\langle u_{1z}^2\rangle < \langle u_{1y}^2\rangle$. Thus, it is
not that ``the vacuum acts as a hard wall", as has been suggested as an
explanation for the reversed anisotropies on Al(110)\cite{aimd}, but
the sub-surface atoms that constitute the hard wall that damps the
vibrations along $z$ of atoms in the surface layer.

Another consequence of the increased value of $\beta_{13}$ is that the
surface will try to always maintain a fixed value for the interlayer
separation $d_{13}$, if necessary at the expense of changes in $d_{12}$
and $d_{23}$. This tendency has been confirmed by additional
calculations in which, upon varying $d_{12}$, $d_{23}$ was found to
change in such a way that $d_{13}$ was approximately constant.  One can
now understand why, upon heating an unreconstructed FCC(110) surface,
while $d_{12}$ and $d_{23}$ may expand/contract, they usually do so in
such a way that $d_{13}$ remains roughly constant, i.e., if one
contracts the other expands.  Of course a full treatment of the thermal
expansion/contraction will require that one take into account the
anharmonicity of the interatomic interactions; however, since the
coefficient of thermal expansion is inversely proportional to the
square of the harmonic force constant, the stiffening of $\beta_{13}$
can be expected to have a big impact on the thermal variation of
$d_{13}$.

The fact that all three materials display the same trends, and the
large impact of allowing for the relaxation of interlayer spacings,
suggest that the enhanced stiffening of $\beta_{13}$ over all other
$\beta_{ij}$s may be more a consequence of the bonding geometry than of
special features of the electronic structure.  There are two relevant
features in the geometry of FCC(110) surfaces: (i) there is a very
large reduction in the co-ordination of surface atoms, as a result of
which the bulk-truncated surface relaxes by decreasing $d_{12}$
significantly,  (ii) a topological peculiarity of FCC(110) surfaces
[but not the (111) or (100) surfaces] is that a surface atom is
connected by NN bonds to atoms in the first, second {\em and third}
layers of atoms parallel to the surface.  This explains why
$\phi_{zz}(1\ 2)$ and $\phi_{zz}(1\ 3)$ may have comparable value, but
not why the latter should be much larger.

However, the large enhancement in $\beta_{13}$ and the smaller increase
in $\beta_{12}$ can be rationalized, {\em a posteriori}, by simple
trigonometry. One has to realize that for the FCC(110) geometry, upon
contraction of $d_{12}$ (and thus also of $d_{13}$), the shortening of
the interatomic bond lengths $r_{12}$ and $r_{13}$  does not scale
uniformly with the contraction of the corresponding interlayer
separations $d_{12}$ and $d_{13}$.  For example, a $10\%$ contraction
in $d_{12}$, relative to the bulk interlayer spacing $d_B$, translates
to a $5\%$ contraction in the NN bond length between atoms in layers 1
and 3, but a contraction of only $2.4\%$ for the NN bond between atoms
in layers 1 and 2.  Since the stiffness of bonds scales as a high power
of their equilibrium length \cite{zepthesis}, this results in a much
larger increase in the radial force constant for 1-3 than 1-2 bonds.

The increase in $d_{23}$ that results from the oscillatory relaxation
of these surfaces weakens this effect. However, these simple geometric
considerations suggest that if $d_{23}$ is still small enough relative
to $d_B$ so that $d_{23}^2 + 2d_{12}d_{23} - 3d_B^2 < 0$, one can still
expect to find that $\beta_{13} \gg \beta_{12}$.  Similarly, if the
contraction of $d_{34}$ is sufficiently small, then the radial force
constant $\beta_{24}$ should be softened, thus increasing $\langle
u_{2z}^2\rangle$ further.

Table I shows the results obtained for selected MSDs using the modified
surface force constants and Equation 1. These results compare well with
those deduced from measurements using low-energy electron diffraction
(LEED), medium energy ion scattering (MEIS) and helium atom scattering
(HAS), as well as MD simulations using either {\em ab initio} (AI-MD)
or Embedded Atom Method (EAM) potentials (there is, however, a
considerable scatter in the values of MSDs available in the
literature).  It is important to note that my results do not include
anharmonic effects, which may be large for FCC (110) surfaces,
especially Cu(110)\cite{anharm}.  Thus any discrepancy between my
results and the experimental or MD ones may indicate that anharmonic
effects are significant.

The calculations presented in this paper can be extended by including
anharmonic interactions within a quasiharmonic approximation, as has
been done for other surfaces \cite{mine,linres}. However, I hope to
have shown that the physics of the harmonic sector is itself
interesting, and can go a long way towards predicting and understanding
the thermal behavior of FCC(110) surfaces.  Given the simplicity of the
geometric arguments, it should also be easy to extend these results to
the unreconstructed (110) surfaces of other FCC metals.


\onecolumn
\begin{figure} \caption{Results from {\em ab initio} calculations for
selected diagonal elements of the interlayer force constant tensors
coupling the first layer of atoms with sub-surface layers n, for the
fully relaxed (110) surfaces of (a) Ag: $xx, yy$ and $zz$ elements (b)
Ag, Al and Cu:  $zz$ elements. The $zz$ elements for bulk Ag are also
shown.} \end{figure}

\begin{table}
\caption{Comparison of calculated and experimental MSDs for atoms in 
the two topmost layers. $T$ is the temperature in Kelvin, and
$\langle u_{bi}^2\rangle$ is the one-dimensional MSD of bulk atoms.  
\label{table1}}
\begin{tabular}{ccccccccc}
Material & Method & $T$(K) & 
$\langle u_{1x}^2\rangle$ & $\langle u_{1y}^2\rangle$ &
                            $\langle u_{1z}^2\rangle$ &
$\langle u_{2x}^2\rangle$ & $\langle u_{2y}^2\rangle$ & 
                            $\langle u_{2z}^2\rangle$  \\
\tableline
Al(110) & this work               & 400 & 0.018 & 0.027 & 0.022 &
                                          0.015 & 0.014 & 0.032 \\
\       & AI-MD\tablenote{Ref. 9} & 400 & \     & 0.028 & 0.019 &
                                          \     & 0.013 & 0.030 \\
\       & LEED\tablenote{Ref. 6}  & 400 & \     & \     & 0.032 & 
                                          \     & \     & 0.030 \\
\       & MEIS\tablenote{Ref. 4}  & 330 & 1.5$\langle u_{bi}^2\rangle$ & 
                                          1.2$\langle u_{bi}^2\rangle$ &
                                          1.1$\langle u_{bi}^2\rangle$ &
                                          \     &     \ & \     \\
\tableline
Ag(110) & this work               & 300 & 0.013 & 0.021 & 0.014 & 
                                          0.010 & 0.010 & 0.026 \\
  \     & EAM\tablenote{Ref. 7}   & 300 & 0.014 & 0.020 & 0.013 &
                                           \    &   \   &   \   \\
  \     & MEIS\tablenote{Ref. 3}  & 300 & 0.022 & 0.048 & 0.026 & 
                                          0.012 & 0.022 & 0.026 \\
\tableline
Cu(110) & this work               & 300 & 0.008 & 0.009 & 0.010 &
                                          0.006 & 0.007 & 0.010 \\
\       & EAM$^{\rm d}$           & 300 & 0.011 & 0.019 & 0.013 &
                                          0.008 & 0.010 & 0.014 \\
\       & HAS\tablenote{Ref. 25}  & 300 &       &       & 0.012 & 
                                                &       &       \\
\end{tabular}
\end{table}



\begin{references}

\bibitem{misrow} M. Copel and T. Gustafsson, Phys. Rev. Lett. {\bf
57}, 723, and references therein.  

\bibitem{oscill} R.N. Barnett {\em et al.}, Phys. Rev. B {\bf 27},
6534 (1983).

\bibitem{agmeis} B.W. Busch and T. Gustafsson, Surf. Sci. {\bf
407}, 7 (1998).

\bibitem{almeis} B.W. Busch and T. Gustafsson, Phys. Rev. B {\bf
61}, 16097 (2000).

\bibitem{alleed1}A. Mikkelsen, J. Jiruse and D.L. Adams, Phys. Rev.
B {\bf 60}, 7796 (1999).

\bibitem{alleed2} H. Goebel and P. von Blanckenhagen, Phys. Rev. B
{\bf 47}, 2378 (1993).

\bibitem{talat} L.~Yang and T.S. Rahman, Phys. Rev. Lett. {\bf 67}
2327 (1991); T.S. Rahman and Z.J. Tian, Journ. of Elec. Spect. and
Relat.  Phenom. {\bf 64/65}, 651 (1993).

\bibitem{nicu110} Y. Beaudet, L.J. Lewis and M. Persson, Phys. Rev.
B {\bf 50} 12084 (1994); H. Hakkinen and M. Manninen, Phys. Rev. B {\bf
46}, 1725 (1992).

\bibitem{aimd} N. Marzari {\em et al.} Phys. Rev. Lett. {\bf 82},
3296 (1999).

\bibitem{stroscio} J.A. Stroscio {\em et al.} Phys. Rev. Lett.
{\bf 54}, 1428 (1985).

\bibitem{persson} M. Persson, J.S. Stroscio and W. Ho, J.
Electron. Spectrosc. Relat. Phenom. {\bf 38}, 11 (1986).

\bibitem{black} J.E. Black {\em et al.} Phys. Rev. B {\bf 36} 2996
(1987).

\bibitem{franchini1}  A. Franchini {\em et al.}, Phys. Rev. B {\bf
38}, 12139 (1988).

\bibitem{handb} K.M.~Ho and K.P. Bohnen, Phys. Rev. Lett. {\bf
56}, 934 (1986).

\bibitem{gaspar} J.A. Gaspar {\em et al.} Phys. Rev. Lett. {\bf
66}, 337 (1991).

\bibitem{franchini2} A. Franchini {\em et al.}, Phys. Rev. B {\bf
47}, 4691 (1993).

\bibitem{ho} Y.~Chen {\it et al.}, Phys. Rev. Lett. {\bf 70}, 603
(1993); S.Y. Tong {\it et al.}, Surf. Rev. and Lett. {\bf 1}, 97
(1994).

\bibitem{helfeyn} H. Hellmann, {\em Einf\"uhrung in die
Quantenchemie} (Deuticke, Liepzig, 1937); R.P. Feynman, Phys. Rev {\bf
56}, 340 (1939).

\bibitem{fhi96md} M. Bockstedte {\em et al.}, Comp. Phys. Comm.
107, 187 (1997).

\bibitem{tmpp} N. Troullier and J.L. Martins, Phys. Rev. B {\bf
43}, 1993 (1991).

\bibitem{cepaldxc} D.M.~Ceperley and B.J.~Alder, Phys. Rev. Lett.
{\bf 45}, 566 (1980).

\bibitem{mine} S. Narasimhan and M. Scheffler, Z. Phys. Chem. {\bf
202}, 253 (1997); S. Narasimhan, Surf. Sci. Lett. {\bf 417}, L1166
(1998).


\bibitem{nilehwald} S. Lehwald {\em et al.}, Surf. Sci. {\bf 131},
 (1987).

\bibitem{msdform} R.E.~Allen and F.W.~de~Wette, Phys.~Rev. {\bf
179}, 873  (1969).

\bibitem{zepthesis} P. Zeppenfeld, Ph.D. Thesis, KFA J\"ulich
(1989).

\bibitem{anharm} P. Zeppenfeld {\em et al.}, Phys. Rev. Lett. {\bf
62}, 63 (1989); G. Armand and P. Zeppenfeld, Phys. Rev. B {\bf 40},
5936 (1989); A.P.~Baddorf and E.W.~Plummer, Phys. Rev. Lett. {\bf 66},
2770 (1991); A.M. Raphuthi {\em et al.}, Phys. Rev. B {\bf52}, R5554
(1995).

\bibitem{linres} J. Xie {\em et al.}, Phys. Rev. B {\bf 59}, 970
(1999).



\end{references}
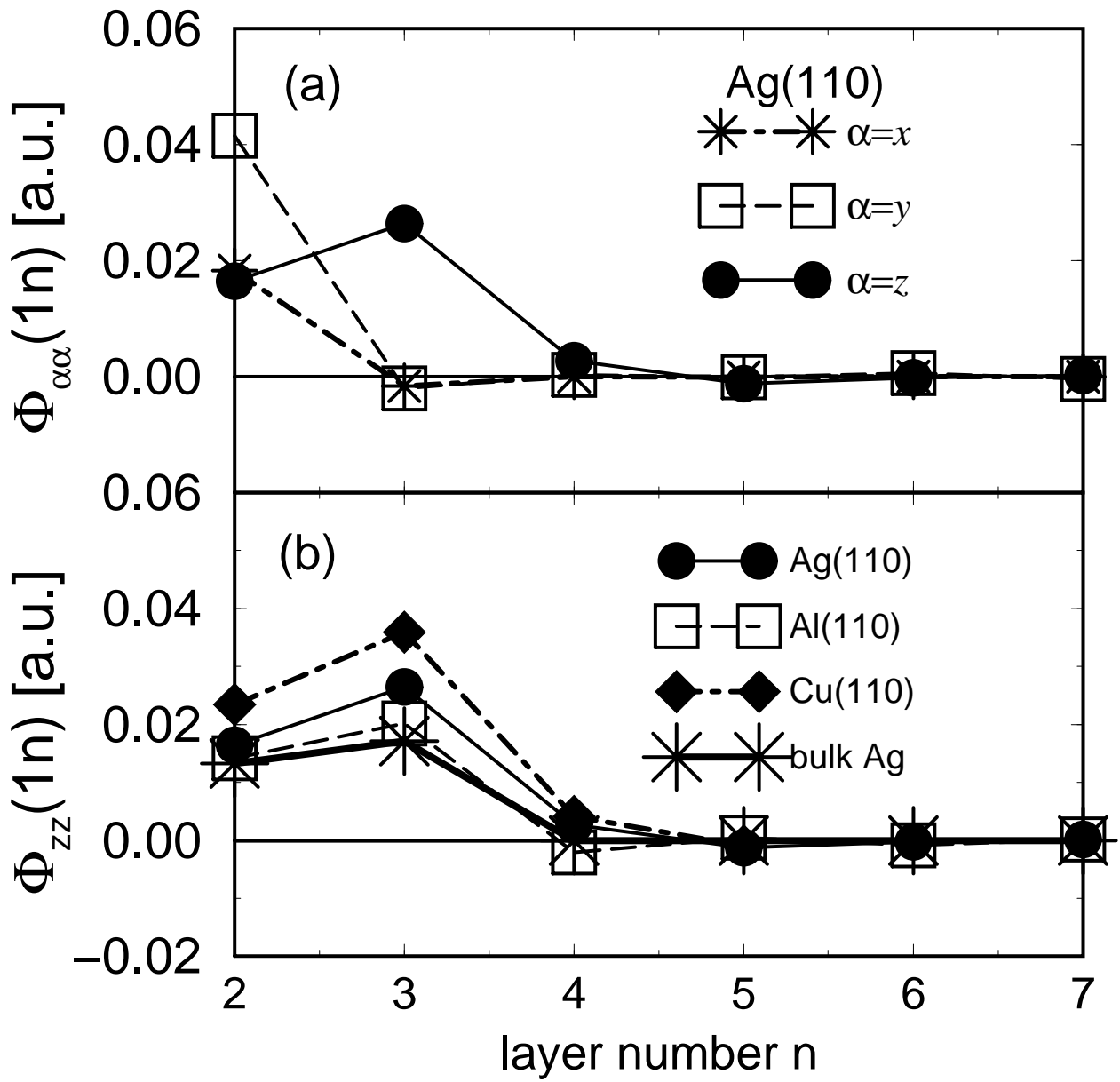
\end{document}